\documentclass[preprint2]{aastex631}
\usepackage{multirow}
\usepackage{threeparttable}

\graphicspath{{./}{}}

\begin{document}

%% TITLE
\title{Environmental impacts on the rest-frame UV size and morphology of
star-forming galaxies at $z\sim2$}

%% AUTHORS
\author[0000-0001-7713-0434]{Abdurrahman Naufal}
\affiliation{Department of Astronomical Science,
The Graduate University for Advanced Studies,
2-21-1 Osawa, Mitaka,
Tokyo 181-8588, Japan}

\author{Yusei Koyama}
\affiliation{Department of Astronomical Science,
The Graduate University for Advanced Studies,
2-21-1 Osawa, Mitaka,
Tokyo 181-8588, Japan}
\affiliation{Subaru Telescope,
National Astronomical Observatory of Japan,
650 North A’ohoku Place,
Hilo, HI 96720, USA}

\author{Rhythm Shimakawa}
\affiliation{Waseda Institute for Advanced Study (WIAS), 
Waseda University, 
1-21-1, Nishi-Waseda, Shinjuku, Tokyo 169-0051, Japan}
\affiliation{Center for Data Science, 
Waseda University, 
1-6-1, Nishi-Waseda, Shinjuku, Tokyo 169-0051, Japan}

\author{Tadayuki Kodama}
\affiliation{Tohoku University,
2-1-1 Katahira, Aoba Ward, Sendai,
Miyagi 980-8577, Japan}

\received{22 February 2023} 
\revised{11 September 2023}
\accepted{18 September 2023}
\submitjournal{ApJ}

%% ABSTRACT
\begin{abstract}

We report the measurement of rest-frame UV size and morphology of H$\alpha$-emission-selected star-forming galaxies (HAEs) in four protoclusters at z $\sim$ 2 (PKS 1138-262, USS 1558-003, PHz G237.0+42.5, and CC 2.2) using archival Hubble Space Telescope Advanced Camera Surveys (HST/ACS) F814W data. We compare the measurement of 122 HAEs in protoclusters detected by HST/ACS to a coeval comparison field sample of 436 HAEs. We find the size distributions of protocluster and field HAEs are similar with typical half-light radius of $\sim$ 2.5 kpc. At fixed stellar mass, there is no significant difference between HAE in protocluster and in field, which is also supported by stacking analyses. This result suggests that the environment does not significantly affect the size of galaxies during the star-forming phase at this epoch. Based on Sérsic index and non-parametric morphologies, HAE morphologies in both environments at $z\sim2$ in rest-frame UV are consistent with disk-like star-forming galaxies, although we also find $29\% \pm 4\%$ HAEs disturbed morphologies. The fraction of disturbed galaxies is higher in protocluster environment, with $39 \pm 8 \%$ protocluster HAEs showing disturbed morphologies, compared to $26\pm4\%$ in the comparison field. The apparent disturbed morphologies are correlated with higher star-formation activity and may be caused by either in situ giant clumps or mergers.

\end{abstract}


%% INTRODUCTION
\section{Introduction} \label{sec:intro}

In the local Universe, we observe the well-known morphology-density relation: galaxy clusters are predominantly populated by the massive, quiescent elliptical and S0 galaxies, while the actively star-forming spiral galaxies are more commonly found in less dense environment \citep{Dressler1980}. This relationship indicates that environment influence the evolution of galaxies by affecting their star formation histories as well as transforming their morphologies.

Local galaxy clusters are thought to be formed from overdense regions in high redshift usually called galaxy protoclusters (for review, see \cite{Overzier2016}). Unlike in local clusters, galaxies in protoclusters are still actively star-forming and assembling the progenitors of the most massive galaxies today \citep[e.g.,][]{Chapman2008, Koyama2013, Dannerbauer2014, Casey2015, Shimakawa2018a, Shimakawa2018b}. In order to become the quiescent-dominated clusters as we see in the local Universe, we might expect that the galaxies in protoclusters to experience accelerated evolution than those in field---such as enhanced star-formation, more rapid quenching, as well as more prevalent morphological transformation due to the dense environment. Some protoclusters are found to already have a higher fraction of quiescent galaxies than their coeval comparison field, supporting the hypothesis of accelerated evolution \citep[e.g.,][]{Kubo2021, Shi2021, McConachie2022SpectroscopicGalaxies}.

The morphology-density relation at higher redshift, however, still remains in question. \citet{Tasca2009} found that the morphology-density relation is already in place at $z = 1$, but the trend is flatter towards higher redshift, indicating morphological evolution of cluster galaxies since $z = 1$. \citet{Newman2014} found that the fraction of disk-like quiescent galaxies in a $z=1.80$ cluster is lower than that in field, implying most of the quiescent galaxies in the cluster have undergone morphological transformation. \citet{Sazonova2020}, using mainly non-parametric morphological analysis methods for clusters at $1.2 < z < 1.8$, found two clusters have more bulge-dominated galaxies than the field, but two other clusters host a population that is indistinguishable from field galaxies. Studying 16 (proto)clusters at $1.2 < z < 1.8$, \citet{Mei2022Morphology-density1.4ltzlt2.8} found that the fraction of spheroid galaxies is higher in clusters, as well as the fraction of mergers. \citet{Peter2007} found the morphologies of UV-selected star-forming galaxies in a protocluster at $z=2.30$ are not different from those in the general field.

Another important measure of the structural evolution of galaxies is the galaxy size. Quiescent or early-type galaxies grow in size by a factor of $\sim4$ \citep[e.g.,][]{Wuyts2011, Newman2012, VanDerWel2014, Shibuya2015} since $z = 2.5$, while star-forming or late-type galaxies grow more slowly \citep[e.g.,][]{VanDerWel2014}. There are likely many mechanisms affecting this growth, some of which are environmentally driven such as mergers, strangulation, or galaxy harassment. At low redshifts, \citet{Cebrian2014TheGalaxies} found that both late-type and early-type galaxies in clusters are smaller than those in field. At $z \sim 1$ clusters, \citet{Allen2016} and \citet{Matharu2019} found that quiescent and star-forming galaxies are smaller than their field counterparts. Towards higher redshifts, some studies show that early-type/quiescent galaxies in clusters are larger than their field counterparts \citep[e.g., ][]{Papovich2012, Lani2013, Allen2015, Andreon2018, Afanasiev2023TheSizes}, although others do not find size difference \citep[e.g.,][]{Newman2014} or even smaller cluster early-type galaxies \citep{Raichoor2012}. Meanwhile, late-type/star-forming galaxies in high-redshift clusters do not show significant difference from their field counterparts \citep[][]{Lani2013, Peter2007, Afanasiev2023TheSizes}, although some studies also find that they are larger \citep[][]{Allen2015, Tran20172}.

The discrepancies of environmental effects on galaxy properties and structures at higher redshifts are not surprising since most high-redshift studies are based on different single (proto)clusters with relatively small sample ($\sim30-60$ galaxies) with various overdensities, sample selection, and redshifts. Ideally, to obtain a more general picture of the environmental effects on galaxy structures at $z\sim2$, one would need to examine the structures of galaxies in various coeval protoclusters, and a uniform sample selection is important in order to obtain a sensible comparison.
%===============================

%%[aim]
In this work, we analyze the size and morphology of star-forming galaxies in protoclusters at $z \sim 2$ and in field. The galaxies in our sample were observed by emission line surveys, e.g., MApping HAlpha and Lines of Oxygen with Subaru \citep[MAHALO-Subaru;][]{Kodama2012} and High-redshift Emission Line Survey \citep[HiZELS; ][]{Geach2012, Sobral2013}. These surveys use narrow-band filters to identify and map the distribution of line-emitting star-forming galaxies at a redshift slice corresponding to the pivot wavelength of the filter used. This enables us to have a large sample of uniformly selected star-forming galaxies at a certain redshift slice---i.e., $z\sim2$ for our work. We compile catalogs of H$\alpha$-emitting galaxies (HAEs) in four protoclusters and in a coeval comparison field that were observed with Hubble Space Telescope Advanced Camera for Survey (HST/ACS), consisting of 122 HAEs in protoclusters plus 446 HAEs as the comparison field sample. With this largest, homogenously H$\alpha$-selected sample of star-forming galaxies at $z\sim2$ with HST/ACS coverage, we examine the size and morphologies in F814W filter, which probes the rest-frame UV in this redshift range. We utilize non-parametric morphology statistics as well as Sérsic profile fitting in order to characterize the size and morphology of HAEs to discern if environment affects the structures of HAEs at this redshift.

We describe our dataset in Section \ref{sec:dataset}. The quantitative morphology diagnostics that we use are briefly described in Section \ref{sec:intro-morph}. The morphological analyses are described in Section \ref{sec:analysis} and the results are presented in Section \ref{sec:results}. We further discuss the results in Section \ref{sec:discussion} and summarize the key points in Section \ref{sec:summary}. Throughout this paper, we adopt $\Lambda$CDM cosmology with $H_0 = 70$ km s$^{-1}$ Mpc$^{-1}$, $\Omega_\mathrm{m} = 0.3$, $\Omega_\mathrm{\Lambda} = 0.7$.

\section{Dataset} \label{sec:dataset}

\subsection{Protoclusters}

We select protoclusters observed as H$\alpha$ overdensities at $z \sim 2$ with available HST/ACS data, as high-resolution image is required to study the morphological properties of high-redshift galaxies. Our HAE sample is compiled from MAHALO-Subaru and HiZELS catalogs. The HAEs are selected using narrow-band filters with pivot wavelengths corresponding to redshifted H$\mathrm{\alpha}$ line wavelength at $z\sim 2$ and a $Ks$ filter for continuum detection. The narrow-band selection criteria are similar, but not identical, for all samples, which imply the sample is H$\mathrm{\alpha}$ flux (and equivalent width) limited. For all samples, the flux limit correspond to a SFR limit down to approximately $3-5 M_\odot \mathrm{yr}^{-1}$. The selected narrow-band emitter samples are then refined by $BzK$ color selection or by photometric redshift to remove contamination. We refer to the reference papers listed in Table \ref{tab:catalog} for the details in the sample selection. 

For each protocluster (and comparison field) HAE, we use the stellar mass and star-formation rate (SFR) derived by the references (see the following sections and Table \ref{tab:catalog}) from SED fitting and H$\alpha$ luminosity, respectively. While the photometric bands used and the methods for SED fitting in each sample are different, we confirm that the stellar masses are distributed similarly (histogram in Figure \ref{fig:mass-sfr}) down to $~10^8 M_\odot$ with comparable uncertainties of $0.1 - 0.3$ dex.

For morphological analyses, we obtain the HST/ACS image of the HAEs from the Hubble Legacy Archive\footnote{\url{https://hla.stsci.edu/hlaview.html}} for PKS 1138-262 and US 1558-003. Since PHz G237.0+42.5, CC 2.2, and the comparison field are located in the COSMOS field, we use v2.0 data processed by \cite{Koekemoer2007} provided by IPAC/IRSA COSMOS cutout service\footnote{\url{https://irsa.ipac.caltech.edu/data/COSMOS/index_cutouts.html}}.

Here we briefly describe each protocluster in our sample as well as the comparison field sample.

\subsubsection{PKS 1138-262}
PKS 1138-262 (or MRC 1138-262; hereafter PKS1138) is a protocluster found around the Spiderweb radio galaxy ($\alpha = 11^\mathrm{h}40^\mathrm{m}48^\mathrm{s}$, $\delta = 26^\mathrm{d}29^\mathrm{m}09^\mathrm{}s$) at redshift $z = 2.16$. The protocluster has been extensively studied since two decades ago, revealing an overdensity of Ly$\alpha$ and H$\alpha$ emitters \citep[e.g.,][]{Pentericci1997, Kurk2000a, Koyama2013} but also a forming red sequence \citep[][]{Kodama2007, Tanaka2010a}. Observations with HST/ACS revealed a complex merging processes occurring in the central galaxy, i.e. a forming brightest cluster galaxy \citep{Miley2006}. As a part of MAHALO-Subaru Deep Cluster Survey (MDCS) campaign, \cite{Shimakawa2018b} identified a total of 68 HAEs; 39 of which are spectroscopically confirmed by \citet{Perez-Martinez2022=2.16}. The survey also finds a high fraction of AGN among massive HAEs \citep[see also ][]{Tozzi2022}. These AGN-hosting HAEs are found to have redder colors, implying that they are in a post-starburst phase. These findings suggest that PKS1138 is a protocluster in maturing phase, where massive galaxies are transitioning from dusty starbursts to quiescent galaxies.

We use HST/ACS F814W data available in the Hubble Legacy Archive (PropID: 10327; PI: Holland Ford, $t_\mathrm{exp} = 4500$ s and $6900$ s), where the protocluster is imaged in two $3'.4 \times 3'.4$ fields. %50 out of 68 HAEs identified by \citep{Shimakawa2018b} is detected in the HST images.

\subsubsection{USS 1558-003}
USS 1558-003, hereafter USS1558, is a protocluster located at redshift $z = 2.53$ ($\alpha = 16^\mathrm{h}01^\mathrm{m}17^\mathrm{s}$, $\delta =-00^\mathrm{d}28^\mathrm{m}47^\mathrm{s}$). It was first discovered by \citet{Kajisawa2006} as an overdensity of red galaxies associated with the radio galaxy USS 1558-003, but was later found to also be populated HAEs spread into three dense groups \citep{Hayashi2012, Hayashi2016}.  In the follow-up campaign MAHALO Deep Cluster Survey, \citet{Shimakawa2018a} identified 107 HAEs tracing four dense groups composing the protocluster within $4' \times 7'$ area and found enhanced star-formation in the densest group. \citet{Macuga2019} finds an AGN fraction of $\sim2\%$ in this protocluster, unlike the relatively higher abundance in PKS1138 and not significantly higher than the field AGN fraction at this redshift. 

We obtain two HST/ACS F814W images covering the USS1558 field from the Hubble Legacy Archive (PropID: 13291; PI: Masao Hayashi, $t_\mathrm{exp} = 5000$ s).

\subsubsection{PHz G237.0+42.5}
PHz G237.0+42.5, hereafter PHzG237, is a protocluster first identified as a Planck compact source at redshift $z = 2.16$ in the COSMOS field ($\alpha = 10^\mathrm{h}01^\mathrm{m}54^\mathrm{s}$, $\delta =02^\mathrm{h}19^\mathrm{m}37^\mathrm{s}$). An overdensity of HAEs was discovered by \citet{Koyama2021} through narrow-band observations with Subaru Multi-Object InfraRed Camera and Spectrograph (MOIRCS). \citet{Koyama2021} identified 38 HAEs in the $4' \times 7'$ observed field, revealing a clustering of massive HAEs at the density peak. \citet{Polletta2021} spectroscopically confirmed 31 member galaxies distributed in two substructures that are expected to collapse into a Virgo-type cluster by $z=0$ and estimated an AGN fraction of $20\pm10\%$.

Since this protocluster is located in the COSMOS field, we obtain the HST/ACS F814W counterparts of the HAEs from IPAC/IRSA COSMOS cutout service\footnote{\url{https://irsa.ipac.caltech.edu/data/COSMOS/index_cutouts.html}} ($t_\mathrm{exp} = 2000$ s).

\subsubsection{CC 2.2}
CC 2.2, hereafter CC22, is a protocluster originally found as an overdensity of the HiZELS sample at $z = 2.23$ ($\alpha = 10^\mathrm{h}00^\mathrm{m}50^\mathrm{s} $, $\delta = 02^\mathrm{d}00^\mathrm{m}02^\mathrm{s}$). \citet{Darvish2020} carried out a follow-up spectroscopic observation with Keck/MOSFIRE near-infrared multiobject spectrograph and confirmed 35 HAE members of the protocluster. Additionally, they found 12 additional galaxies observed by zCOSMOS-Deep Survey in the vicinity of the protocluster, potentially also associated with the protocluster. The protocluster has a higher fraction of X-ray detected HAE AGN than the HiZELS COSMOS field by a factor of $\sim 4$. Since both CC22 and the field sample are from HiZELS COSMOS field, we use spectroscopically confirmed HAEs in CC22 in our work to avoid field contamination.

As a part of the COSMOS field, we obtain the HST/ACS F814W counterparts of the HAEs from IPAC/IRSA COSMOS cutout service ($t_\mathrm{exp} = 2000$ s).

\subsection{Comparison field}

We use HAEs in the HiZELS $z = 2.23$ catalog \citep[][]{Sobral2014}, excluding 35 galaxies that are identified as a member of CC 2.2, as our coeval comparison field. There are 5 galaxies in HiZELS sample that lies within $5'$ from the center (HAE density peak) of CC 2.2, but not included in the spectroscopic member catalog. We decide not to include the 5 galaxies in either sample. We obtain the HST/ACS F814W counterparts of the HAEs from IPAC/IRSA COSMOS cutout service ($t_\mathrm{exp} = 2000$ s).

\subsection{Detection and catalog matching}

For our targets covered by COSMOS, we rely on the HST/ACS COSMOS catalog published by \citet{Leauthaud2007}. We match the HAE catalogs to the ACS catalog with a separation criterion $r < 1"$. We use the IRSA/IPAC COSMOS cutout service to obtain a $5" \times 5"$ cutout for each object detected by ACS. We choose to obtain the cutouts from COSMOS tile images instead of the mosaicked image, because the mosaicked one does not provide a corresponding weight image for each cutout. The pixel scale of the COSMOS images is $0".03/\mathrm{pix}$ and a PSF with full-width at half-maximum of $\sim 0.11"$.

For our non-COSMOS targets, i.e. PKS1138 and USS1558, we first obtain their HST/ACS F814W images from the Hubble Legacy Archive (HLA). HLA provides 4 and 2 tile images along with the corresponding weight images for PKS1138 and USS1558, respectively, at a pixel scale of $0".05/\mathrm{pix}$. We produce a source catalog by running SExtractor following the same procedures and configuration parameters as in \citet{Leauthaud2007} for each image, using the provided weight image as the input for WEIGHT\_IMAGE parameter. Similar to the COSMOS sample, we match the HAE catalogs to the source catalogs with separation $r < 1"$. We created a $5" \times 5"$ cutout for each detected objects using Astropy package\footnote{https://www.astropy.org}.

With separation $r < 1"$, we find that 184 field HAEs and 54 protocluster HAEs have more than one ACS-detected objects within the radius centered on the HAE coordinates. In this case, we select the nearest object to the HAE coordinate listed in the catalog. From only F814W data, we cannot confirm or rule out whether the additional objects are associated with the HAEs or not. In the following morphological properties measurements, nearby separate objects are masked.

Since the HAE catalogs are compiled from ground-based emission line observations, some HAEs in the catalogs are not covered by their respective HST/ACS F814W observations, are located on the edge of the tile images, or are undetected in this filter. After these steps, we identified the HST/ACS F814W counterparts of 467 field HAEs and 148 protocluster HAEs.

\begin{table*}[t]
\begin{center}

\caption{Summary of the protocluster and field HAE sample, which describes, for each sample: the protocluster redshift $z$, the number of HAEs in the reference catalog $N_\mathrm{HAE}$, the number of HAEs detected in HST/ACS $N$, $5\sigma$ point source depth in $0".25$ diameter apertures corrected by Milky Way extinction, the PSF FWHM based on stacked stars, the number of unflagged non-parametric morphology measurements $N_\mathrm{good}$, and the number of unflagged Sérsic fitting results $N_\mathrm{Sersic}$.}

\begin{threeparttable}
\begin{tabular}{lcccccccc}
\hline
\multicolumn{1}{c}{\multirow{2}{*}{Name}} & \multirow{2}{*}{$z$} & \multirow{2}{*}{Ref} & \multirow{2}{*}{$N_{HAE}$} & \multicolumn{5}{c}{F814W}                                                  \\ \cline{5-9} 
\multicolumn{1}{c}{}                      &                      &                      &                          & $N$ & 5$\sigma$ (ABmag)\tnote{a} & FWHM (") & $N_\mathrm{good}$ & $N_\mathrm{Sersic}$ \\ \hline
PKS 1138-262                              & 2.16                 & \tnote{1}                    & 68                       & 35  & 28.0            & 0.10     & 28                & 28                  \\
USS 1558-003                              & 2.53                 & \tnote{2}                    & 107                      & 28  & 27.2            & 0.12     & 10                & 20                  \\
PHz G237.0+42.5                           & 2.16                 & \tnote{3}                    & 38                       & 27  & 27.4            & 0.11     & 14                & 27                  \\
CC2.2                                     & 2.23                 & \tnote{4, 5}                 & 35                       & 32  & 27.5            & 0.11     & 17                & 25                  \\ \hline
\textit{Protoclusters}   & \multicolumn{1}{l}{} &                      & 248                      & 122 &                 &          & 69                & 98                  \\
Field                                     & 2.23                 & \tnote{4}                    & 553                      & 436 & 27.3            & 0.11     & 258               & 404                 \\ \hline
\end{tabular}

\begin{tablenotes}\footnotesize
\item[a] Corrected by Galactic extinction (\url{http://irsa.ipac.caltech.edu/applications/DUST/}).
\item[1] \citet{Shimakawa2018b}
\item[2] \citet{Shimakawa2018a}
\item[3] \citet{Koyama2021}
\item[4] \citet{Sobral2014}
\item[5] \citet{Darvish2020}
\end{tablenotes}

\end{threeparttable}

\label{tab:catalog}
\end{center}
\end{table*}

\subsection{Final sample} \label{sec:final-sample}

\begin{figure}
    \centering
    \includegraphics{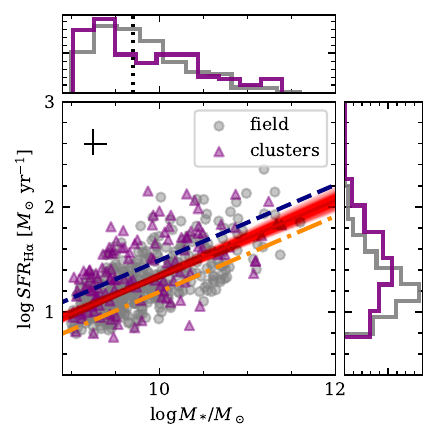}
    \caption{The SFR-stellar mass diagram for our full sample. The red line is a linear fit to field sample only to estimate the  star-forming main sequence (SFMS). The average uncertainties in SFR and stellar mass are shown in top left corner. Above (below) the blue (orange) line is the `above (below) SFMS' sample that is be used in our analyses. The black dashed line in the horizontal histogram marks the division between low-mass and high-mass sample. Protocluster and field HAEs are distributed similarly in this diagram to illustrate that we are comparing a similar population of galaxies in two different environments.}
    \label{fig:mass-sfr}
\end{figure}

\begin{figure}
    \centering
    \includegraphics[scale=0.6]{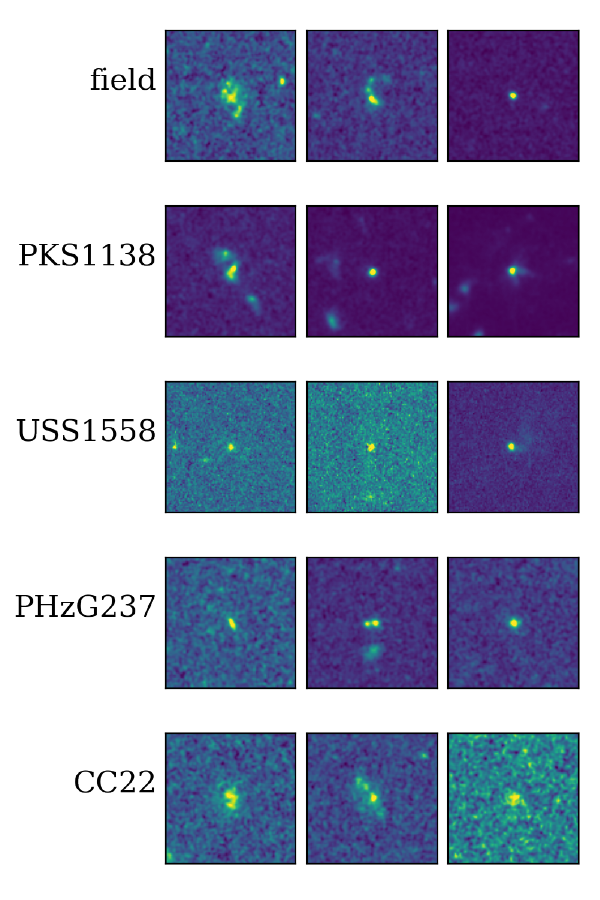}
    \caption{For each sample, we show three randomly-selected example HST/ACS F814W $5"\times5"$ cutouts ordered by increasing mass from left to right. We see various apparent morphologies: compact spheroid-like, disky, and irregular or disturbed.}
    \label{fig:cutouts}
\end{figure}

We find that our protocluster HAE sample has a lower median stellar mass than our field sample, $10^{9.49} < 10^{9.69} \; \mathrm{M_\odot}$, mainly due to the more low-mass HAEs in the USS1558 catalog, although Anderson-Darling test shows that the two samples are likely drawn from the same parent distribution ($p=0.10$). Since we expect that galaxy size is correlated with stellar mass (the mass-size relation), we chose to apply simple a stellar mass cut mitigate for stellar mass bias, $\log M_\star / \mathrm{M_\odot} > 9$, which approximately equalize the medians to $10^{9.70} \; \mathrm{M_\odot}$. We do not include objects flagged as X-ray sources in the original catalogs, i.e., 13/248 objects in protoclusters and $10/553$ objects in field.

Finally, our sample consists of 558 H$\alpha$-selected star-forming galaxies at redshift $z = 2.16-2.53$, 122 of which are in protoclusters (for brevity, this sample is referred as 'clusters' in the figures). The number of HAEs after the final selection is shown in Table \ref{tab:catalog}.

The distribution of stellar masses and star-formation rates (SFR) is shown in Figure \ref{fig:mass-sfr}, showing that the HAE samples selected from two kinds of environment are similarly distributed. Thus, we are comparing two populations of similarly selected star-forming galaxies residing in different environments. We also show some HAE cutout examples in Figure \ref{fig:cutouts}.

%%%%%%
\section{Quantitative morphology diagnostics} \label{sec:intro-morph}
The study of galaxy structures and morphologies started by the means of visual classification \citep[][]{Hubble1926ExtragalacticNebulae.}. This method is not effective for distant galaxies, as we are limited by image resolution. We also cannot assume that high-redshift galaxies with similar morphologies as local galaxies have the same physical properties (for example, assuming early-type galaxies [ETG] are quiescent or late-type galaxies [LTG] are star-forming at all redshifts)---in fact, investigating the correlation between galaxy structures and physical properties is the motivation of the study of galaxy morphologies. Therefore, we use quantitative morphology indicators to investigate the variation in galaxy morphology and examine these correlations more objectively.

There are two approaches to quantify the morphological properties of galaxies: parametric approach and non-parametric approach. In the parametric approach, an assumed light profile model is fitted to the data. In the case of galaxies, Sérsic profile is commonly used \citep[]{Sersic1963}, which generalizes the form described by \citet{deVaucouleurs1948RecherchesExtragalactiques}. Non-parametric approaches are designed to capture the features of the underlying structures of galaxies without making assumptions about the underlying form. In this section, we will describe some quantitative morphology indicators that will be used in this work.

\subsection{Sérsic profile}

\citet{deVaucouleurs1948RecherchesExtragalactiques} proposed a fitting form for radial light profile of elliptical galaxies. This form was then generalized by \citet{Sersic1963} to also describe different types of galaxies,

\begin{equation}
    I(R)=I_{0} \times \exp \left\{-b(n) \times\left[\left(R / R_{\mathrm{e}}\right)^{1 / n}-1\right]\right\},
\end{equation}

where the profile is described by Sérsic index $n$ and $R_\mathrm{e}$ is the effective radius containing half of the light contained within the galaxy. The classic de Vaucouleurs profile has $n = 4$ that describes elliptical galaxies, while exponential disks are described by $n = 1$. Some works fit a galaxy with two Sérsic profiles, decomposing the profile of a galaxy into a bulge component and a disk component.

\subsection{Concentration} \label{sec:concentration}

The concentration index \citep[]{Bershady2000, Conselice2003} quantifies how concentrated the light of a galaxy is by comparing the light in the inner parts of the galaxy to that in the outer parts. It is commonly defined as the circular radius containing $80\%$ of the light ($r_{80}$) to the radius containing $20\%$ of the light ($r_{20}$),

\begin{equation}
    C=5 \log \left(\frac{r_{80}}{r_{20}}\right).
\end{equation}

The total flux is usually taken as the light contained within $1.5$ Petrosian radius, where Petrosian radius is the radius where the surface brightness is $20\%$ of the mean surface brightness.

Elliptical galaxies and bulge-dominated spirals have high concentration values of $C \gtrsim 4$, while galaxies with no significant bulge will have low concentration values.

\subsection{Asymmetry}

The asymmetry index \citep[][]{Conselice2000} measures how asymmetric a galaxy is by measuring the difference of a galaxy image ($I_{(x, y)}$) with its $180^\circ$-rotated image ($I_{180(x, y)}$) corrected by the average asymmetry of the background $B_{180}$,

\begin{equation}
    A=\frac{\sum_{x, y}\left|I_{(x, y)}-I_{180(x, y)}\right|}{2 \sum\left|I_{(x, y)}\right|}-B_{180}.
\end{equation}
The center of rotation is initially guessed by the physical center, but then iteratively determined to minimize the value. Asymmetry value may be negative if the local background is high enough and the surface brightness of the galaxy is low.

Elliptical galaxies generally have low asymmetry indices due to their lack of structures. Local spiral galaxies generally have $A = 0.05 - 0.20$ \citep[][]{Conselice2014}. Irregular galaxies and merging galaxies have high asymmetry values due to clumpy or lop-sided morphologies. For $z > 1$ galaxies, \citet{Lotz2008} proposed the criterion $A > 0.30$ to capture candidates of merging galaxies. \citet{Lotz2010a, lotz2010b} also found that asymmetry can capture gas-rich mergers with baryonic mass-ratio down to 1:4.

Another parameter that is often used with Concentration and Asymmetry is the Smoothness (or Clumpiness) parameter. Smoothness attempts to quantify the fraction of light in a galaxy that is contained in clumps by taling the difference between the galaxy image and a blurred galaxy image. However, this parameter is not used in our work as we find that it is not effective in detecting barely-resolved clumps of high-redshift galaxies.

\subsection{Gini-M$_{20}$} \label{sec:gini-m20}

Gini coefficient, adapted from economics, measures the distribution of light among the pixels that comprise a galaxy \citep[][]{Lotz2004}. The pixels are sorted with increasing value, and the coefficient is determined by

\begin{equation}
    G=\frac{1}{\bar{X} n(n-1)} \sum_{i}^{n}(2 i-n-1) X_{i},
\end{equation}
where $n$ is the number of pixels in the galaxy’s segmentation map, $X_i$ is the pixel flux at the rank $i$ pixel and $\bar{X}$ is the mean pixel value. Gini segmentation map is constructed according to the Petrosian radius, which is therefore insensitive to the $(1 + z)^4$ surface brightness dimming of distant galaxies \citep{Lotz2004}. This segmentation map is created convolving the galaxy image with a Gaussian kernel with $\sigma = r_\mathrm{petro}/5$, where $r_\mathrm{petro}$ is the elliptical Petrosian semimajor axis. The mean surface brightness at $r_{petro}$ is used to define a flux threshold, so that pixels with flux values above this threshold are assigned to the galaxy.

As such, an object with uniformly distributed light will have Gini coefficient of $0$, while an object concentrated to a single pixel has a value of $1$. Gini coefficient is correlated to concentration when the galaxy is centrally concentrated. However, Gini coefficient can also be high when there is off-center bright regions in the galaxy.

Another coefficient often paired with Gini is the second-order moment M$_{20}$. M$_{20}$ measures the second-order moment of the $20\%$ brightest pixels of the galaxy. First, the total second-order moment is measured by multiplying the flux in each pixel
$fi$ by the squared distance to the center of the galaxy, summed over all the galaxy pixels assigned by the segmentation map,
\begin{equation}
    M_{\mathrm{tot}}=\sum_{i}^{n} M_{i}=\sum_{i}^{n} f_{i}\left[\left(x_{i}-x_{c}\right)^{2}+\left(y_{i}-y_{c}\right)^{2}\right],
\end{equation}
where $xc$, $yc$ is the galaxy center that is iteratively computed to minimize $M_{tot}$. M$_{20}$ is then computed by ranking the galaxy pixels by flux, summing $Mi$ over the brightest pixels until the sum of the brightest pixels equals $20\%$ of the total galaxy flux, and then normalize by $M_{tot}$,

\begin{equation}
    M_{20}=\log \left(\frac{\sum_{i} M_{i}}{M_{\mathrm{tot}}}\right), \text{while} \sum_{i} f_{i}<0.2 f_{\text {tot }}.
\end{equation}

M$_{20}$ complements Gini to differentiate galaxies with bright off-center clumps from those with bulges. Values for M$_{20}$ are generally between $-0.5$ and $-2.5$. Elliptical galaxies generally have lower (more negative) values while spiral and irregulars have higher values. Studying galaxies in $0.2 < z < 0.4$, \citet{Lotz2008} classified galaxies with $G>-0.14 M_{20}+0.33$ as merging galaxies. They also found that non-merging galaxies with $G>0.14 M_{20}+0.80$ are ellipticals and lenticulars.

Expanding on those classification, Gini-M$_{20}$ is used to define 'bulge statistic' \citep[][]{Snyder2015a, Snyder2015b, Rodriguez-Gomez2019}:
\begin{equation}
    F\left(G, M_{20}\right)=-0.693 M_{20}+4.95 G-3.96,
\end{equation}
and 'merger statistic':
\begin{equation}
    S\left(G, M_{20}\right)=0.139 M_{20}+0.990 G-0.327.
\end{equation}.

\citet{Lotz2008} found that the merger classification by Gini and M$_{20}$ can capture mergers down to mass ratio of 1:9 independent of the gas fractions in the galaxies.

%%%%%%
\section{Analysis} \label{sec:analysis}

We study the morphological properties of HAEs in our sample using the $\textsc{Python}$ package \textsc{Statmorph} v0.4 \citep{Rodriguez-Gomez2019}. \textsc{Statmorph} is an \textsc{Astropy}-affiliated package for calculating mainly non-parametric morphology statistics of galaxies from images. The morphology statistics calculated by \textsc{Statmorph}  include Concentration, Asymmetry and Smoothness (CAS) indices \citep{Conselice2014, Lotz2004}; Gini-M20 coefficients \citep{Lotz2004, Snyder2015a, Snyder2015b}; and several more along with shape and size measurements associated to the above statistics (ellipticity, Petrosian radius, half-light radius, MID statistics, shape asymmetry, etc.). In addition to non-parametric properties, \textsc{Statmorph} can also fit the parametric Sérsic profiles to galaxy images \citep{Sersic1963}.

We performed size and morphology statistics measurements on stacked images of HAEs as well as individual HAE cutouts, as described below.

\subsection{\textsc{Statmorph} configuration} \label{sec:statmorph}

\textsc{Statmorph} requires an input image, an input segmentation image, and gain parameter or a weight map as its minimum input. As the input image, we use a $5" \times 5"$ cutout for each input object centered at F814W detection, rescaled to $0.05" \mathrm{pix}^{-1}$ for COSMOS images.

The segmentation image is mainly used to mask other sources in the FOV of the target. In case of two nearby target source, for example, \textsc{Statmorph} will measure one target while masking the other, and then do the same for the other one's turn. We use the package \textsc{Photutils} to create a segmentation image of sources with $1.5\sigma$ detection above the background. A segmentation image generated by \textsc{SExtractor} can also be used, but since we do not have the segmentation image generated for the COSMOS ACS sources, we decided to treat all sample uniformly by using \textsc{Photutils} segmentation images. By visually comparing \textsc{SExtractor}-generated segmentation images and \textsc{Photutils} segmentation images for our non-COSMOS sources, we confirm that \textsc{Photutils} can reasonably reproduce \textsc{SExtractor} segmentation images.

The weight map that \textsc{Statmorph} expects is a sigma image with the same size and units as the input image. In case a weight map is not provided, the gain parameter is required to generate a weight map internally. Note that \textsc{Statmorph} expects gain as a multiplication factor that converts the image units into electrons per pixel. Since HST/ACS images are provided in electrons per second per pixel, the exposure time is appropriate for this parameter. HST/ACS images provided by the Hubble Legacy Archive and the COSMOS Cutouts Service include the corresponding weight images in the form of inverse variance map, i.e., $w = 1/\sigma^2$.

A PSF image can also be input to \textsc{Statmorph} to improve the Sérsic model fitting. The proposed model would be convolved with the PSF first before being compared to the data. We use empirical PSF for each sample derived from stacking point sources in the images. The PSF image is not used by non-parametric morphology statistics.

We ran \textsc{Statmorph} for all detected HAEs in our sample. \textsc{Statmorph} flag some measurements as unreliable due to reasons such as inadequate skyboxes, low S/N, or failed Sérsic fitting. These flagged objects are excluded in our results (see Section \ref{sec:results}).

\begin{figure*}
    \centering
    \includegraphics[scale=1]{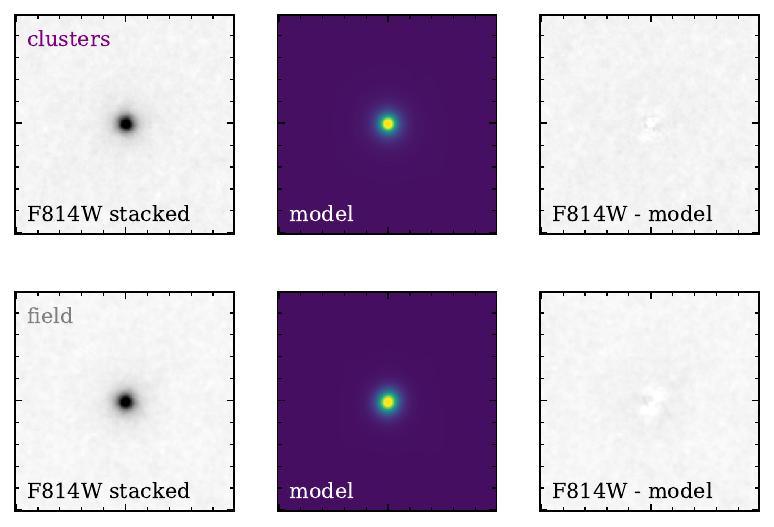}
    \caption{Each row shows an example of a stacked image (left) and the Sérsic model (middle). Right panels show the residuals of the from the image. Note that these stacked images are one instance of the bootstrap sampling (Section \ref{sec:stacking}).}
    \label{fig:stacking}
\end{figure*}

\begin{figure}
    \centering
    \includegraphics[scale=0.5]{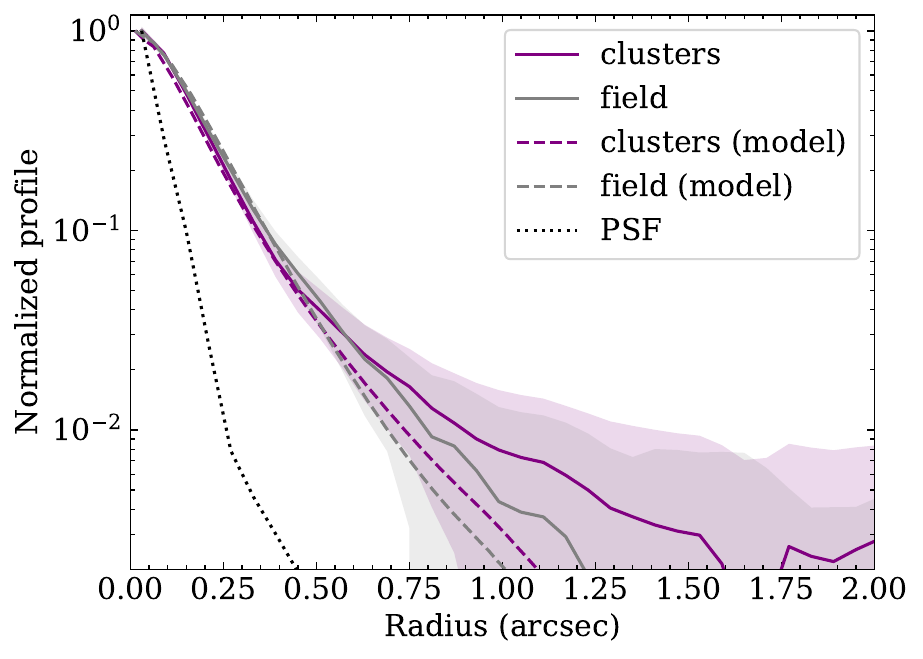}
    \caption{Radial profiles of stacked field HAEs (solid gray line) and stacked protocluster HAEs (solid purple line). The shaded regions are uncertainties of the stacked profile from bootstrapping.  The radial profiles of best-fitting Sérsic models are shown as dashed lines and the PSF is shown as the black dotted line.}
    \label{fig:radialprofile}
\end{figure}

\subsection{Stacking procedure} \label{sec:stacking}

We stacked HST/ACS F814W images of HAEs in order to measure the average size of HAEs in our sample. By analyzing the stacked galaxy, we can derive the typical size of each group without being biased by the brightest sample. We use all ACS-detected sample for the stacking analysis, regardless of their individual measurement flags.

The procedures for stacking is as follows:
\begin{enumerate}
    \item We create $5" \times 5"$ cutouts for each object centered at F814W detection. Since COSMOS images have a smaller pixel scale than non-COSMOS images, we opt to resize COSMOS cutouts to match the pixel scale of non-COSMOS cutouts, i.e., from $0".03 \mathrm{pix}^{-1}$ to $0".05 \mathrm{pix}^{-1}$.

    \item Each image is smoothed with the corresponding PSF matching kernel to the largest FWHM of observational PSF in our sample of $0".128$, which is the PSF of USS1558 images.

    \item We randomly select the same number of galaxies from field sample and from protocluster sample for the stacking.

    \item Each cutout is normalized by the total flux of the object and stacked by median flux per pixel centered at the F814W detection without rotating the cutout to align the object's orientation (see section \ref{sec:hae-size}).

\end{enumerate}

The stacked image is then input to \textsc{Statmorph} with similar procedures as in the previous section. We use $2 \times \mathrm{gain} \times N / 3$ as the effective gain, where $N$ is the number of input images. Since the stacking procedure increases the effective S/N but erases detail of each object such as ellipticities, clumpiness, or asymmetry, the main property to be derived from a stacked galaxy image is the size. We performed bootstrapping in step (3) 1000 times and report the median values. We take the 16\textsuperscript{th} and 84\textsuperscript{th} of the distribution as the uncertainties in the stacked measurement.

%In order to estimate the measurement uncertainties of stacked galaxies, we first estimate the 2D background of each (resized) cutout and performed similar procedures. The stacked background RMS is used to generate a random Gaussian noise to be applied to the original image. We repeat the measurement procedure 1000 times and take the median as the average value, with 16\textsuperscript{th} and 84\textsuperscript{th} of the distribution as the uncertainties of the average. The stacked images, best-fitting Sérsic models, and residuals are shown in Figure \ref{fig:stacking}.

\section{Results} \label{sec:results}

\subsection{Typical size of HAEs} \label{sec:hae-size}

\textsc{Statmorph} calculates the half-light radius of an object with several methods: one by parametric approach with Sérsic profile and two non-parametric approaches. The parametric fitting of Sérsic profile takes into account the PSF, while the non-parametric approaches do not.  Hereafter we present the Sérsic half-light radius, $r_\mathrm{Sersic}$, as our main result for the aforementioned reason.

We succesfully fit Sérsic profiles for 502/558 HAEs in our sample. Excluding flagged Sérsic measurements, we find that protocluster HAEs have median size of $2.50^{+0.19}_{-0.19} \; \mathrm{kpc}$, while field HAEs have median of $2.57^{+0.12}_{-0.05} \; \mathrm{kpc}$. The uncertainties in each sample median are the 16\textsuperscript{th} and 84\textsuperscript{th} of the bootstrapped distribution medians. From the stacking analysis, we obtain a median size of $2.23^{+0.22}_{-0.16} \; \mathrm{kpc}$ for stacked protocluster HAEs and $2.21^{+0.14}_{-0.17} \; \mathrm{kpc}$ for stacked field HAEs. While the values of stacked measurement differ from individual measurement, we find that protocluster HAEs are similar to field HAEs in both cases. However, we also note that the stacked image for protoclusters shows more extended outskirts, although still within $1\sigma$, as can be seen in Figure \ref{fig:radialprofile} despite the similar Sérsic effective radii, which may hint to the presence of disturbances or faint companions in protocluster galaxies.

The difference between stacked galaxies and individual galaxies is partly due to the different orientation angles. We do not align the orientation angles of individual galaxies before the stacking due to the large uncertainty from the irregular morphologies. To estimate how this uncertainty affect the size measurement, we follow \citet{Allen2017ZFOURGE} by creating synthetic cutout images of galaxies with uniform size and profile according to the averages from our individual measurement result. We stack the galaxies in two ways: by allowing random orientations and by aligning the major axes. We find that the stacked galaxies with random orientations give $\sim24 \%$ smaller sizes than those with aligned orientation angles. Meanwhile, both methods agree on the Sérsic index values, i.e., randomly-oriented stacking gives a Sérsic index value within $1\sigma$ of aligned stacking.

%This discrepancy is also related to the circularized effective radius that is sometimes used to report galaxy sizes \citep[e.g.,][]{Shen2003}, where $r_\mathrm{circ} = r_\mathrm{Sersic} \sqrt{b/a}$ and $b/a$ is the ratio of the projected axes and $r_\mathrm{Sersic}$ is the semimajor axis as measured by Sérsic profile fitting. If we consider the circularized radii of individual galaxies instead, we have $r_\mathrm{Sersic, \: circ} = 1.83_{-0.71}^{+1.57} \; \mathrm{kpc}$ for protocluster HAEs and $r_\mathrm{Sersic, \: circ} = 1.83_{-0.77}^{+1.12} \; \mathrm{kpc}$ for field HAEs, which are more consistent with our stacking results. In the following discussions, we will keep presenting the semimajor axes $r_\mathrm{Sersic}$ as `size' or `half-light radii' to keep consistency with previous studies.

\subsection{Mass-size relation} \label{sec:mass-size}

\begin{figure*}
    \centering
    \includegraphics{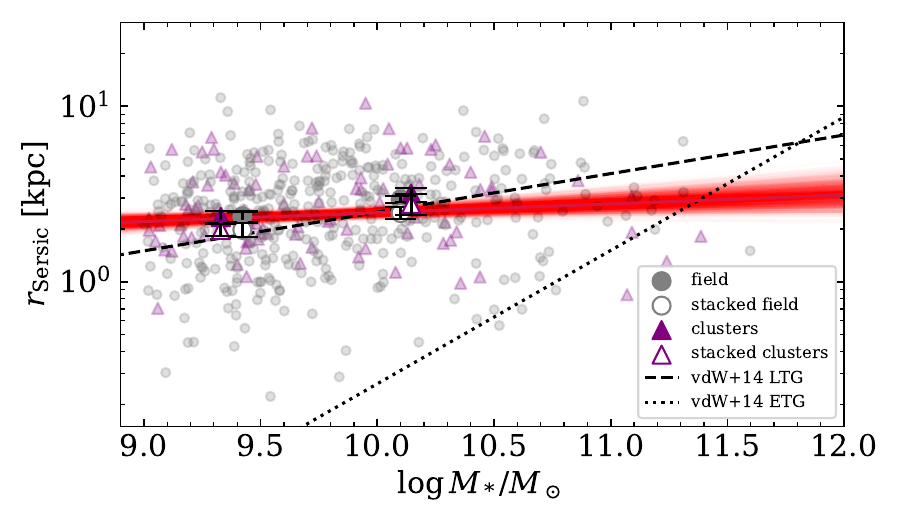}
    \caption{The stellar mass--half-light radius relation of HAEs in our sample. Semi-transparent purple (gray) triangles (circles) are HAEs in protoclusters (field), while their solid counterparts are binned according to the stellar mass. White-filled triangles (circles) are the stacked galaxy of the same mass bin for protocluster (field) HAEs, multiplied by the correction factor (see text). The red line is the derived fit for field galaxies (Equation \ref{eq:mass-size}), to compare with the results from \citet{VanDerWel2014} in dashed and dotted lines for LTG and ETG, respectively. We do not find a significant difference between protoclusters and field samples in either mass bin.}
    \label{fig:mass-size}
\end{figure*}

\begin{figure}
    \centering
    \includegraphics{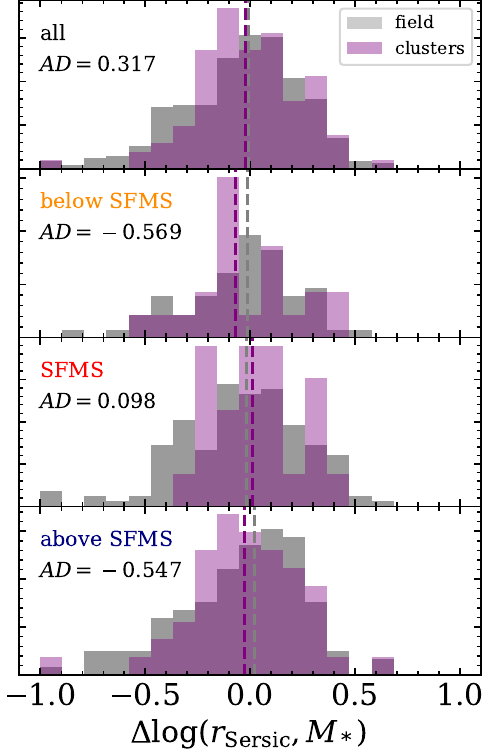}
    \caption{The distribution of $\log{r_\mathrm{Sersic}}$ to the one predicted by Eq. \ref{eq:mass-size}. We divide the samples into three SFR bins as shown in the bottom three panels. In each panel, the distribution for the protocluster (field) sample is represented by purple (gray) histogram. Each panel includes galaxies within $\mathrm{SFR}-M_\star$ ranges as illustrated in Figure \ref{fig:mass-sfr}. For all subsets, we do not find significant difference of size at fixed stellar mass according to Anderson-Darling tests.}
    \label{fig:size-offset}
\end{figure}

\begin{table*}[]
\centering
\bgroup
\def\arraystretch{1.5}
\caption{\textbf{The size for each sample as measured by Sérsic fitting. The stacked $r_\mathrm{Sersic}$ are listed here after multiplying by correction factor of 1.24 (see text). The values in parentheses are the number of samples in each group.}}
\begin{tabular}{cccc}
\hline
                   & \multicolumn{3}{c}{$r_\mathrm{Sersic}$ {[}kpc{]}}                                 \\ \cline{2-4} 
                   & All                       & $\log M_* / M_\odot < 9.7$ & $\log M_* / M_\odot > 9.7$ \\ \hline
Clusters (stacked) & $2.23^{+0.22}_{-0.16}$ (122) & $1.98^{+0.17}_{-0.15}$ (62)  & $2.77^{+0.59}_{-0.38}$ (60) \\
Field (stacked)    & $2.21^{+0.14}_{-0.17}$ (436) & $1.98^{+0.20}_{-0.18}$ (217) & $2.46^{+0.24}_{-0.19}$ (219) \\ \hline
Clusters (binned)  & $2.50^{+0.19}_{-0.19}$ (98)  & $2.34^{+0.20}_{-0.18}$ (52)  & $3.17^{+0.34}_{-0.18}$ (46) \\
Field (binned)     & $2.57^{+0.12}_{-0.05}$ (404) & $2.36^{+0.13}_{-0.14}$ (208) & $2.91^{+0.11}_{-0.21}$ (196) \\ \hline
\end{tabular}
\egroup
\label{tab:mass-size}

\end{table*}

\begin{table*}[]
\centering
\bgroup
\def\arraystretch{1.5}
\caption{\textbf{Similar to Table \ref{tab:mass-size}, but for Sérsic index $n$}}
\begin{tabular}{cccc}
\hline
                   & \multicolumn{3}{c}{$n$}                                 \\ \cline{2-4} 
                   & All                       & $\log M_* / M_\odot < 9.7$ & $\log M_* / M_\odot > 9.7$ \\ \hline
Clusters (stacked) & $1.96^{+0.23}_{-0.29}$ (122) & $1.62^{+0.12}_{-0.12}$ (62)  & $2.31^{+0.39}_{-0.51}$ (60) \\
Field (stacked)    & $1.62^{+0.12}_{-0.12}$ (436) & $1.44^{+0.14}_{-0.14}$ (217) & $1.73^{+0.19}_{-0.18}$ (219) \\ \hline
Clusters (binned)  & $2.50^{+0.19}_{-0.19}$ (98)  & $1.05^{+0.14}_{-0.14}$ (52)  & $0.79^{+0.29}_{-0.20}$ (46) \\
Field (binned)     & $2.57^{+0.12}_{-0.05}$ (404) & $0.75^{+0.07}_{-0.06}$ (208) & $0.78^{+0.13}_{-0.02}$ (196) \\ \hline
\end{tabular}
\egroup
\label{tab:sersic-index}

\end{table*}

We divide our sample into two mass bins: a low-mass subsample $\log M_\star / M_\odot <9.7$ and a high-mass subsample $\log M_\star / \mathrm{M_\odot} > 9.7$. The number of bins is chosen so that we have the same number of galaxies in each bin for the stacking procedure. We carried out the stacking procedures and size measurement for each subsample. We plot the half-light radius $r_\mathrm{Sersic}$ against the stellar masses in Figure \ref{fig:mass-size}, where we can see a mild mass-size relation (derived from field galaxies only). The values are reported in Table \ref{tab:mass-size}. We find that protocluster HAEs are smaller than field HAEs in each mass bin except for the high-mass bin of binned individual measurements, though still within the uncertainties in the case of binned individual measurements.

We parametrize the relation following \citet{VanDerWel2014}:
\begin{equation} \label{eq:mass-size}
    \log r_\mathrm{Sersic} = \log A + \alpha \log{ \left(\frac{M_\star}{5 \times 10^{10} \; M_\odot}\right) }\; \mathrm{kpc} .
\end{equation}
and fit the line using Monte-Carlo Markov Chain (MCMC) method.

We find the slope $\alpha = 0.05 \pm 0.02$ (red line in Fig. \ref{fig:mass-size}), which is shallower than that derived in \cite{VanDerWel2014} (dashed black line). It is likely that this difference is due to the different probed rest-frame wavelengths, as \cite{VanDerWel2014} probes rest-optical wavelengths as opposed to rest-frame near-UV in our work, and the different sample selection. Our result is consistent with \cite{Paulino-Afonso2017}, who also use the HiZELS sample and HST/ACS F814W data and found the slope $\alpha = 0.05 \pm 0.11$ for $z=2.23$ . The $y$-intercept of our fitting gives the average size of $2.63_{-0.15}^{+0.17} \; \mathrm{kpc}$ at $M_\star=5 \times 10^{10} \; \mathrm{M_\odot}$, slightly smaller than that derived in \cite{VanDerWel2014}. We also divide the protocluster sample and field sample into two subsamples each according to their position relative to the star-forming main sequence: `below SFMS' and `above SFMS' for logarithmic offsets larger than $0.2 \mathrm{dex}$ below and above the main sequence, respectively. We do not see any clear trend in either groups.

We calculate the offset between galaxy size and the predicted size at given stellar mass, $\Delta \log(r_\mathrm{Sersic}, M_\star)$, in order to check whether subtle size difference at fixed stellar mass. Protocluster HAEs are only $0.01$ dex ($\sim 2 \%$) smaller on average than field HAEs and we cannot rule out that both samples are drawn from the same parent distribution according to Anderson-Darling test ($p > 0.05$). In Figure \ref{fig:size-offset}, we compare the distributions of size offsets for HAEs below, on, and above SFMS (as illustrated in Figure \ref{fig:mass-sfr}). While we find that HAEs above the main sequence are $0.04$ dex smaller than the typical size at fixed stellar mass, it is also not statistically significant Anderson-Darling test. For each subsets, we cannot rule out that the two distributions are from the same parent distribution. This result suggests that protocluster environment play little role in affecting the size of star-forming galaxies.

\subsection{Morphological properties of HAEs} \label{sec:morphologies}

\begin{figure*}
    \centering
    \includegraphics[width=\textwidth]{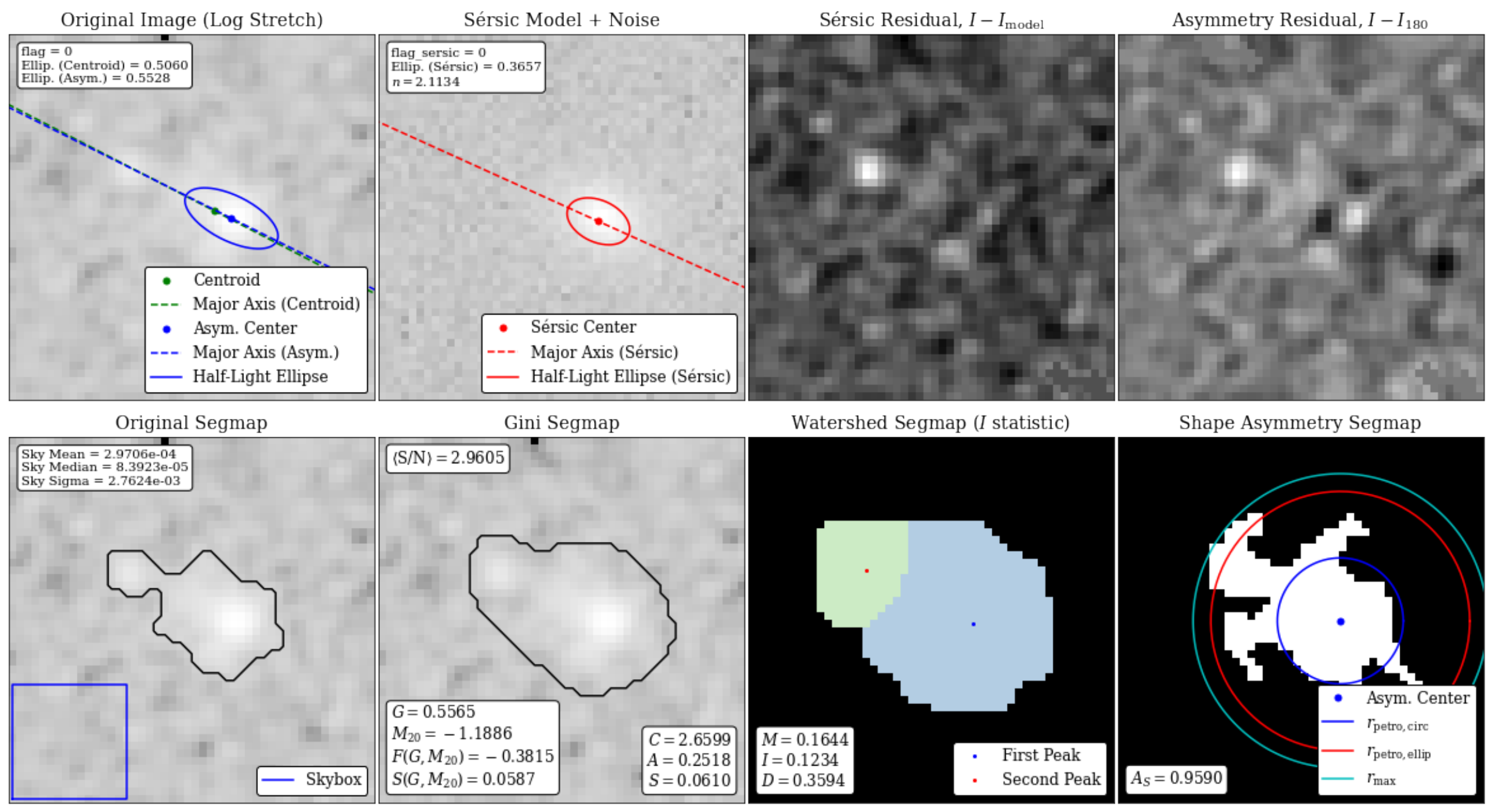}
    \caption{An example figure produced by \textsc{STATMORPH} of morphological measurement of an HAE in CC22. Properties relevant to Sérsic fitting, asymmetry, Gini, M$_{20}$, MID, and shape asymmetry are shown. The two bottom right panels, showing MID and shape asymmetry, are not discussed in this work.}
    \label{fig:statmorph}
\end{figure*}

\begin{figure*}
    \centering
    \includegraphics{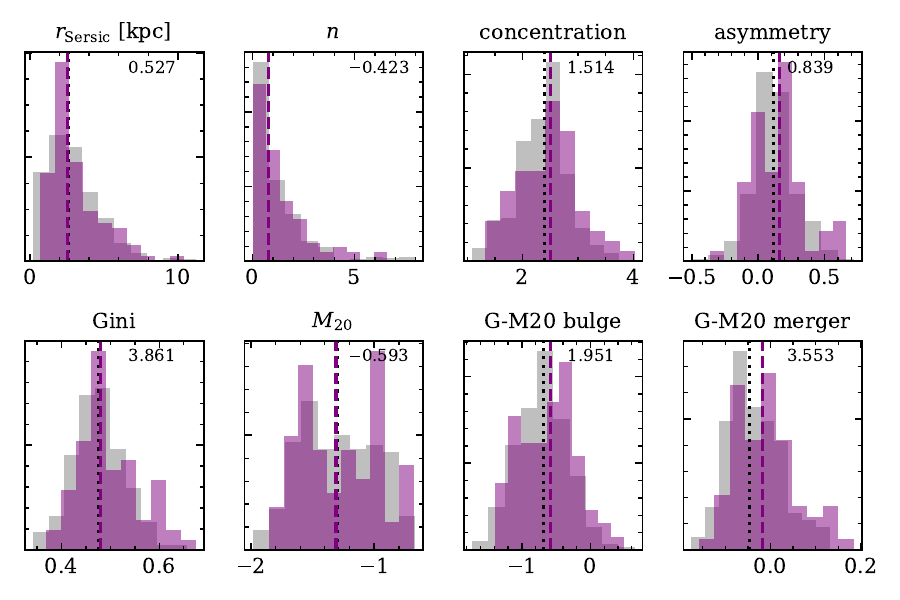}
    \caption{Histogram of morphology indicators of protocluster (purple) and field (gray) HAEs. From top left to bottom right: Sérsic half-light radius, Sérsic index, concentration, asymmetry, Gini, M$_{20}$, and two statistics derived from Gini-M$_{20}$: bulge factor and merger statistic. Purple dashed (black dotted) vertical lines denote the median of protocluster (field) sample.} In each panel we show the Anderson-Darling coefficients, where $AD > 1.961$ corresponds to $p < 0.05$ significance. We find significant difference in Gini and the two Gini-M$_{20}$ statistics.
    \label{fig:morphologies}
\end{figure*}

\begin{figure}
    \centering
    \includegraphics{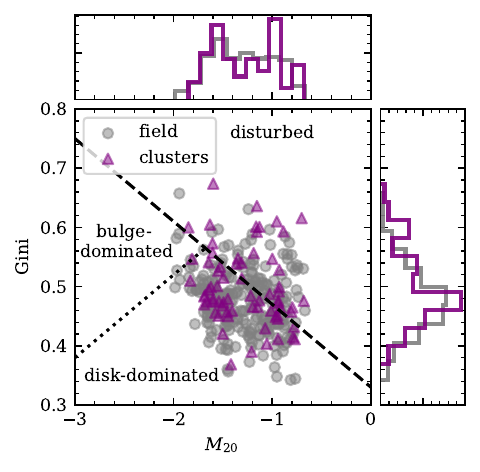}
    \caption{A diagram showing the protocluster HAEs (purple triangles) and field HAEs (gray circles) in the Gini-M$_{20}$ plane. The dotted line is the zero line of the bulge statistics as it segregates disk-dominated (below, negative bulge statistics) and bulge-dominated (above, positive bulge statistics galaxies. Similarly, the dashed line is the zero line of merger statistics above which the merger candidates lie (positive merger statistics) \citep{Snyder2015b, Rodriguez-Gomez2019}. As seen here, most HAEs in our sample are consistent with disk-dominated profile, but also a significant fraction of them has peculiar morphologies that possibly indicate clumpy or merging systems.}
    \label{fig:gini-m20}
\end{figure}

Next, we investigate whether there is any subtler structural difference between protocluster HAEs and field HAEs by characterizing their apparent morphologies. We find that \textsc{Statmorph} failed at measuring the non-parametric morphology statistics of a fraction of our sample. Only 69/122 protocluster and 258/412 field HAEs are deemed as having reliable non-parametric measurement by \textsc{Statmorph}. Most of the flagged objects are due to disjoint Gini segmentation map, which may be caused by the low surface brightness or the low S/N of galaxies in our sample. An example of STATMORPH result is shown in Figure \ref{fig:statmorph}.

The distributions of $r_\mathrm{Sersic}$ and Sersic $n$ from Sérsic fitting as well as non-parametric concentration, asymmetry, Gini, and M$_{20}$ are shown in Figure \ref{fig:morphologies}. The median values in each statistic are shown by the vertical lines. The last two panels show `bulge statistic' and `merger statistic' derived from Gini and M$_{20}$ \citep[][; see Section \ref{sec:gini-m20}]{Snyder2015a, Snyder2015b, Rodriguez-Gomez2019}. The Gini-M$_{20}$ plane is shown in Figure \ref{fig:gini-m20} with demarcation lines that distinguish (nearby) galaxy morphologies according to bulge statistic and merger statistic. The medians of Sérsic index is also summarized in Table \ref{tab:sersic-index}.

%The two statistics derived from Gini-M$_{20}$ is shown in the last two panels of Figure \ref{fig:morphologies}, as described in Section \ref{sec:intro-morph}. Positive Gini-M$_{20}$ bulge factor indicates the object is likely to be bulge-dominated. Gini-M$_{20}$ merger statistic represent how likely it is for a galaxy to be undergoing mergers according to the observed light distribution. Positive values for this statistic lie above the dashed line in Figure \ref{fig:gini-m20}. As an example, two merging high-redshift galaxies may appear as an object consisting of two nuclei. This object will have a positive Gini-M$_{20}$ merger statistics. However, we cannot really distinguish a merging galaxies or a galaxy with prominent clumps solely on our imaging data, and hereafter such objects are referred to as `disturbed galaxies'.

%Our results show that the bulge factor from Gini-M$_{20}$ agree with Sérsic index and concentration. 

According to Sérsic index and Gini-M$_{20}$, we find the majority of HAEs in our sample have disk-like morphologies. We find that protocluster HAEs on average, have slightly higher values of concentration and bulge statistics, as well as higher Sérsic index in stacked images. This may indicate that protocluster HAEs have slightly more concentrated light profiles than field HAEs, although the medians of Sérsic index from individual measurement for both samples are similar, i.e., $n\sim 0.7$. Based on two-sample Anderson-Darling tests, protocluster and field HAEs are unlikely ($AD > 1.961$ for $p < 0.05$) to be drawn from the same parent distributions for Gini coefficient and the statistics derived from Gini-M$_{20}$. The $p$-values of all the comparison tests are given in Table \ref{tab:anderson-darling}. However, they are still consistent with disk-dominated profiles ($n \sim 1$ and bulge factor $< 0$). We find that $29 \pm 4\%$ ($14\pm3\%$) HAEs in combined sample have merger statistic $> 0$ (asymmetry $> 0.3$), indicating disturbed apparent morphologies are also common, which may be caused by the presence of clumps or disturbance in the galaxies. This is also likely the cause of the discrepancy between Sérsic indices from stacked measurement and individual measurement (see Table \ref{tab:sersic-index}), since stacking increases the S/N and smooths out the disturbances in individual galaxies.

\begin{figure*}[]
    \centering
    \includegraphics{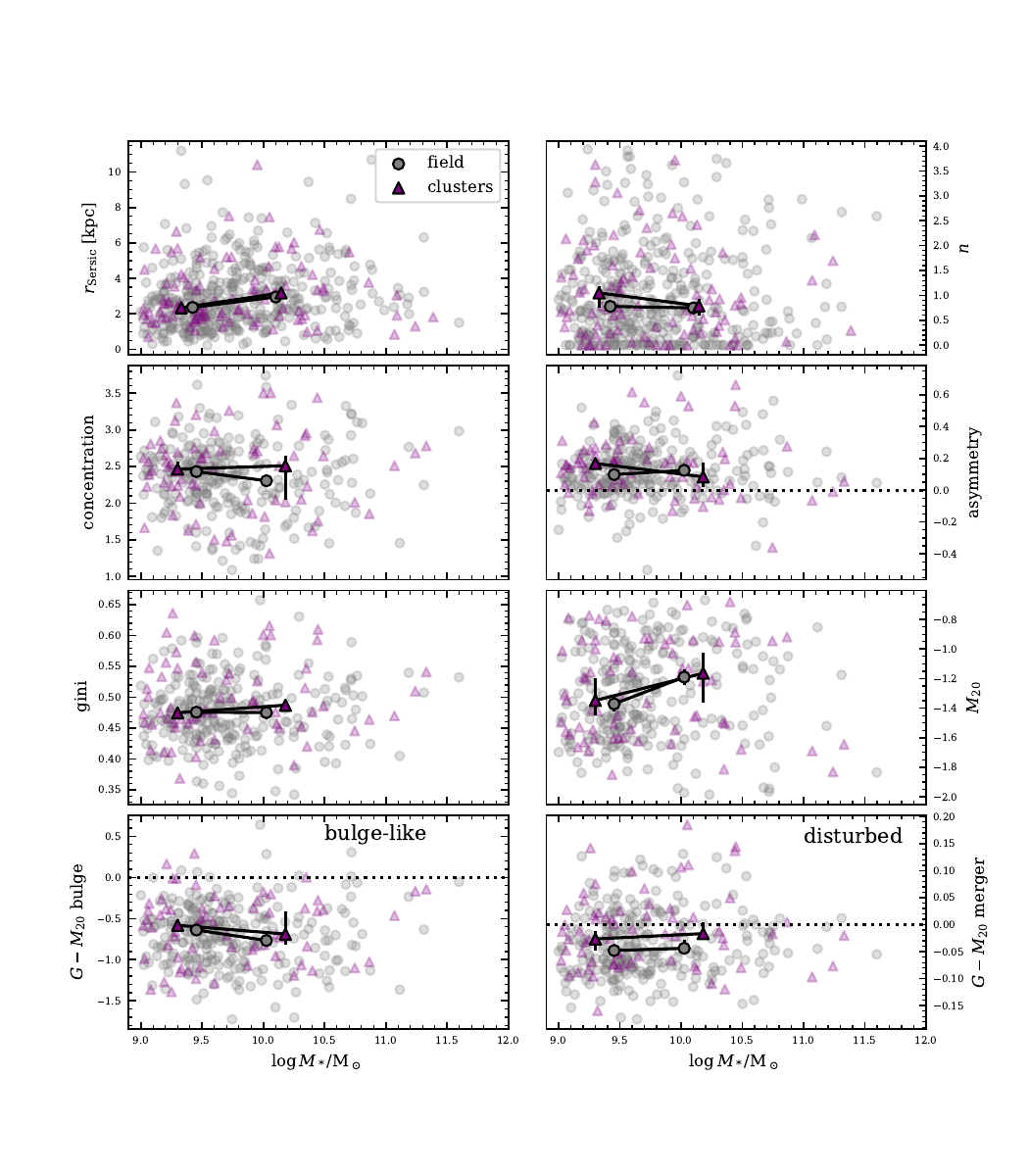}
    \caption{Each panel shows a plot of a morphology statistic against stellar mass $\log M_\star / \mathrm{M_\odot}$. The stellar mass bins are the same as in Figure \ref{fig:mass-size}. The black-bordered symbol represents the median of each bin. The errorbars are the 16th and 84th percentile of the bootstrapped median distribution}. Only stellar mass-size relation (top left) is statistically significant correlation according to Spearman test ($p<0.05$).
    \label{fig:morph-by-mass}
\end{figure*}

\begin{figure*}
    \centering
    \includegraphics{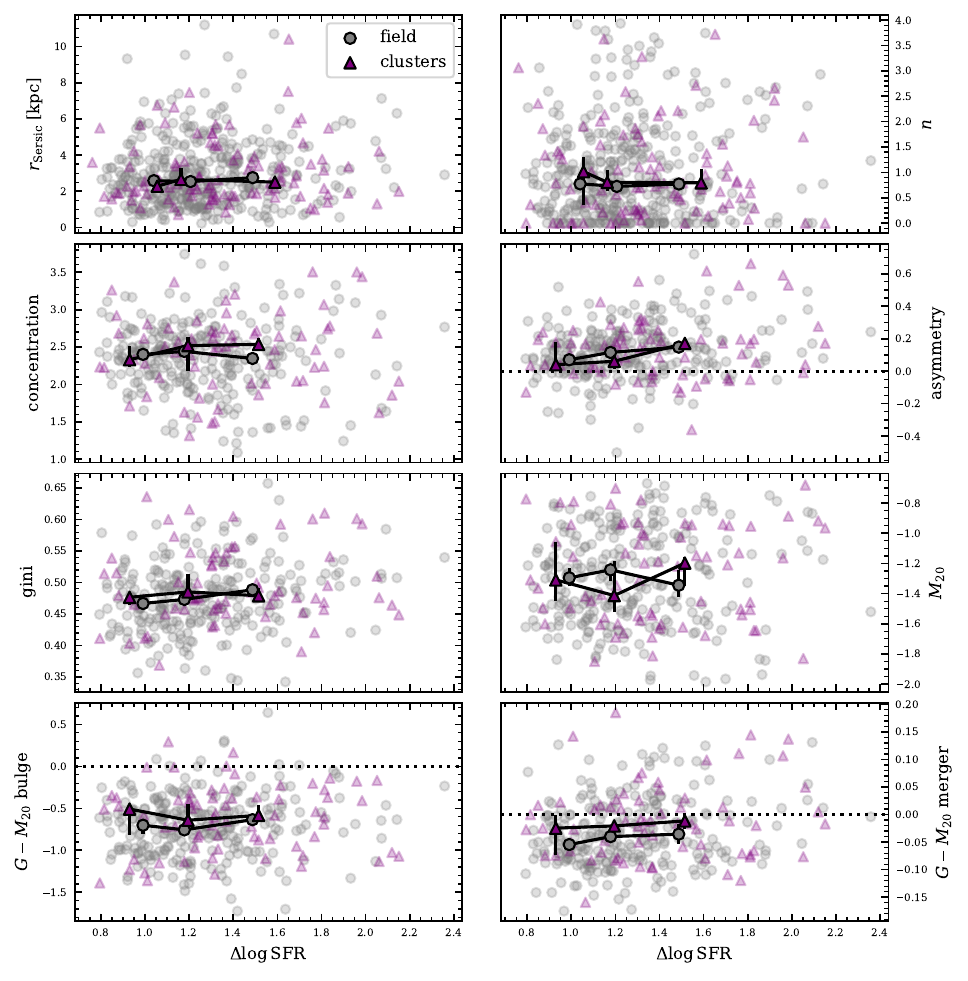}
    \caption{Similar to Figure \ref{fig:morph-by-mass}, but each panel shows a plot of a morphology statistic against offset from SFMS $\Delta\log SFR$. The three bins corresponds to `below SFMS', `on SFMS', and `above SFMS' samples. Asymmetry, Gini, and merger statistics are positively correlated with $\Delta\log SFR$ ($p<0.05$).}
    \label{fig:morph-by-sfr-offset}
\end{figure*}

In Figures \ref{fig:morph-by-mass} and \ref{fig:morph-by-sfr-offset}, we plot morphological properties against stellar mass and SFMS offset, respectively. We divide the distribution of each physical properties into three bins and calculate the median in each bin. For each relationship, we calculate the Spearman's rank correlation coefficient $\rho$. We only find a statistically significant correlation (probability of no correlation $p < 0.05$) between $r_\mathrm{Sersic}$ and stellar mass for field galaxies, which is the mass-size relation as we discussed in Section \ref{sec:mass-size}.

Between morphological properties and SFMS offset (Figure \ref{fig:morph-by-sfr-offset}), we find a positive correlation with asymmetry for the entire samples with Spearman's coefficient $\rho = 0.21$ ($p = 0.002$). Gini and merger statistics are also correlated with the offset from SFMS with $\rho = 0.18$ ($p = 0.01$) and $\rho = 0.16$ ($p = 0.001$), respectively. The correlation between $\Delta \log{\mathrm{SFR}}$ and asymmetry/merger statistics indicates that the highly star-forming galaxies tend to be disturbed in the (rest-frame UV) morphology.

%%%%%%%
\section{Discussion} \label{sec:discussion}

\subsection{Size of star-forming galaxies at $z \sim 2$} \label{sec:scatter}

We find the typical size of HAEs at $z \sim 2$ to be $\sim 2.5 \; \mathrm{kpc}$. This is consistent with \citet{Paulino-Afonso2017} who also used HiZELS catalog and HST/ACS F814W data. At stellar mass $M_\star=5 \times 10^{10} \; \mathrm{M_\odot}$, we find the average size of $2.63_{-0.17}^{+0.16} \; \mathrm{kpc}$, around $0.2$ dex smaller than that derived in \citet{VanDerWel2014}, likely due to the difference in probed wavelength. Note that their star-forming galaxy sample selection is based on color-color diagram while ours is based on H$\alpha$ emission.

As we have also presented in the previous section, we find that size distributions of HAE in protoclusters and field are similar. On the top panel of Figure \ref{fig:size-offset}, we see that there is no significant difference in size between protocluster HAEs and field HAEs at fixed stellar mass.

\begin{figure*}
    \centering
    \includegraphics{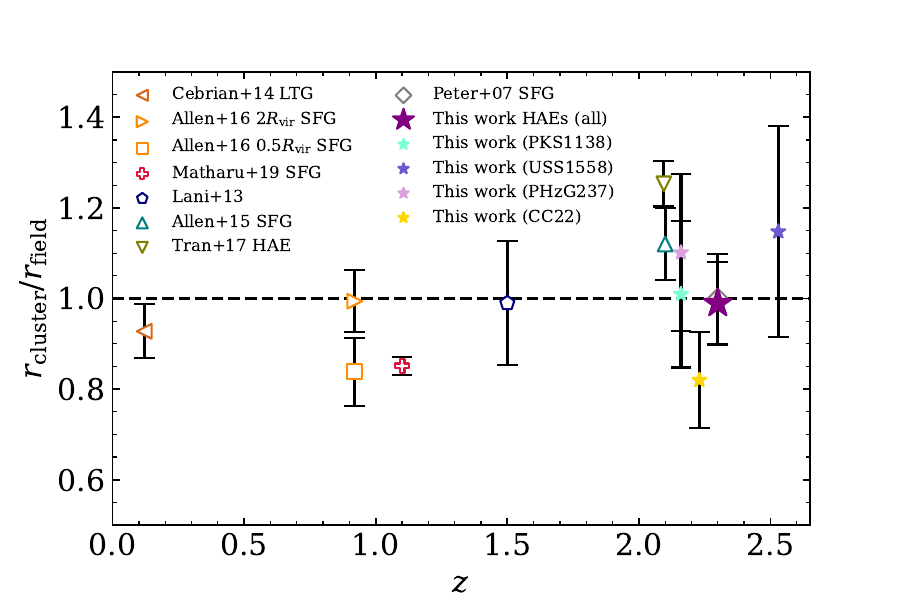}
    \caption{This plot summarizes various studies on star-forming galaxies in different environments \citep{Cebrian2014TheGalaxies, Allen2016, Matharu2019, Lani2013, Allen2015, Tran20172, Peter2007}. $r_\mathrm{cluster}/r_\mathrm{field}$ is the ratio of the size of star-forming galaxies in clusters to those in field at fixed stellar mass (if such value is unreported, we use the ratio of the median of the distributions). Our work provides four data points at $z\sim 2$ (cyan, pink, gold, and slate blue stars for individual protocluster; purple star for all protoclusters). We do not find a significant difference of size at fixed stellar mass between protocluster and field HAEs at $z\sim 2$.}
    \label{fig:prev_works}
\end{figure*}

We compare our result with the previous works at various redshifts in Figure \ref{fig:prev_works}. The lack of size difference for star-forming galaxies is consistent with the findings from \citet{Peter2007} as well as from \citet{Lani2013} at $1<z<2$, but at odds with the findings from \citet{Allen2015}, which found that the mean size at stellar mass $M_\star=5 \times 10^{10} \; \mathrm{M_\odot}$ of star-forming galaxies in a $z = 2.1$ cluster is $12\%$ larger than that of field galaxies. Meanwhile, \citet{Tran20172} did not find any difference between the size distributions of cluster H$\alpha$-selected star-forming galaxies and those in the field, but when looking size distributions at fixed stellar mass, they found that cluster star-forming galaxies are $\sim 0.1$ dex larger than field star-forming galaxies, in line with \citet{Allen2015}. More recently, \citet{Xu2023AcceleratedZ=2.51} found that star-forming galaxies in a $z=2.51$ protocluster is smaller on average than those in field, particularly among the massive sample.

\begin{figure}
    \centering
    \includegraphics{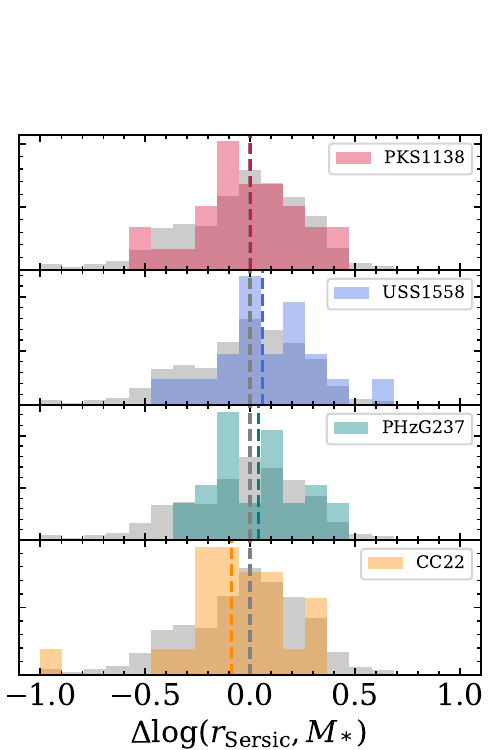}
    \caption{Similar to Figure \ref{fig:size-offset}, but for each cluster compared to field sample in gray histograms. We find some variations in the median offsets, but we cannot rule out that each distribution is drawn from the same parent distribution as the field sample.}
    \label{fig:cluster_variations}
\end{figure}

In Figure \ref{fig:cluster_variations}, we show four panels similar to Figure \ref{fig:size-offset} but for each cluster. When we look at individual protoclusters, we see that CC22 has a slightly lower size offset median than the comparison field, while USS1558 and PHzG237 have higher medians. This fact indicates that there is a variation between each protocluster, which may be caused by differences in their own evolutionary stages. Nonetheless, Anderson-Darling test between each protocluster sample and the comparison field sample shows that none of them are unlikely to be drawn from a different parent distribution as the field sample. Thus, we do not find a strong evidence of accelerated size growth of HAEs at this redshift range. This could mean that star-forming galaxies in clusters and field grow in similar way, until $z < 1$ where they are affected by environmental effects that inhibit size growth in nearby clusters \citep[e.g., ][]{Cebrian2014TheGalaxies, Matharu2019, Matharu2021HST/WFC3Maps}. In contrast, quiescent/early-type galaxies in clusters are found to be larger than their field counterparts at $z > 1.5$ \citep[e.g., ][]{Papovich2012, Andreon2018}, which suggests accelerated size growth. %This difference shows that quenched galaxies grow via different pathway from star-forming galaxies.

\subsection{Environmental dependence of morphology} \label{sec:morphologies-env}

\begin{table*}[]
\centering
\caption{This table lists the median value of each properties for protocluster sample and field sample and Anderson-Darling $k$-sample test $p$-value between the two samples. We chose $p$-value threshold of $0.05$, which corresponds to Anderson-Darling critical value $AD=1.961$, over which the null hypothesis that the two samples are drawn from the same parent distribution cannot be ruled out.}
\begin{tabular}{lccc}
\hline
\multicolumn{1}{c}{\multirow{2}{*}{Parameter}} & \multicolumn{2}{c}{Median}                                  & AD $p$-value         \\
\multicolumn{1}{c}{}                           & Field                        & Protoclusters                & \multicolumn{1}{l}{} \\ \hline
$\log M_\star$                                      & ${9.71}_{-0.03}^{+0.03}$     & ${9.68}_{-0.10}^{+0.10}$     & $0.10$              \\
$\log SFR$                                            & ${1.23}_{-0.01}^{+0.01}$     & ${1.37}_{-0.04}^{+0.03}$     & $0.001$              \\ \hline
\multicolumn{4}{l}{Sérsic fitting (unflagged only)}                                                                                 \\
$r_\mathrm{Sersic}$                            & $2.58^{+0.14}_{-0.05}$       & $2.54^{+0.22}_{-0.21}$       & $0.20$               \\
Sérsic $n$                                     & ${0.80}_{-0.04}^{+0.22}$     & ${0.76}_{-0.05}^{+0.03}$     & $>0.25$              \\
$\Delta\log{(r_\mathrm{Sersic}, M_\star)}$     &                              &                              &                      \\
--all                                          &                              &                              & $>0.25$              \\
--below SFMS                                   &                              &                              & $>0.25$              \\
--on SFMS                                      &                              &                              & $>0.25$              \\
--above SFMS                                   &                              &                              & $>0.25$              \\ \hline
\multicolumn{4}{l}{Non-parametric statistics (unflagged only)}                                                                      \\
Concentration                                  & ${2.40}_{-0.02}^{+0.03}$     & ${2.51}_{-0.06}^{+0.06}$     & $0.08$               \\
Asymmetry                                      & ${0.12}_{-0.03}^{+0.01}$     & ${0.16}_{-0.05}^{+0.01}$     & $0.15$               \\
Gini                                           & ${0.48}_{-0.00}^{+0.00}$     & ${0.48}_{-0.00}^{+0.01}$    & $0.01$               \\
M$_{20}$                                       & ${-1.30}_{-0.05}^{+0.05}$    & ${-1.31}_{-0.10}^{+0.12}$    & $>0.25$              \\
Bulge statistics                               & ${-0.68}_{-0.05}^{+0.03}$    & ${-0.58}_{-0.08}^{+0.08}$    & $0.05$               \\
Merger statistics                              & ${-0.04}_{-0.00}^{+0.01}$    & ${-0.02}_{-0.01}^{+0.01}$    & $0.01$              \\ \hline
\label{tab:anderson-darling}
\end{tabular}
\end{table*}

In Section \ref{sec:morphologies} we presented that HAEs at $z \sim 2$ are consistent with disk-dominated profiles according to quantitative morphological analysis. The majority of HAEs have Sérsic indices $n \sim 1$, consistent with exponential disks. This is also supported by the stacking analysis, which allows us to probe the underlying light profile of HAEs by blurring out the clumps, by finding the Sérsic indices of $1.55^{+0.33}_{-0.29}$ and $1.02^{+0.09}_{-0.08}$ for protocluster and field, respectively. The higher Sérsic index of protocluster HAEs agree with the slightly higher median values of concentration and Gini-M$_{20}$ bulge statistic, although by lesser significance. These trends indicate that protocluster HAEs are slightly more centrally concentrated than field HAEs. There is a possibility that our HAE sample is still contaminated by AGN undetected in X-ray. Since AGN may be more common in protocluster environment \citep[e.g., ][]{Shimakawa2018b}, the measured morphology may be affected. If we include the X-ray sources in the measurement, we find that the differences of Sérsic index, concentration, and bulge statistic are more significant (Anderson Darling test $p < 0.05$), although the size distributions are still undistinguishable.

The higher median merger statistic of the protocluster sample implies that protocluster HAEs are more peculiar/clumpy. Comparing to previous works, \citet{Sazonova2020}, who investigated the morphology-density relations in four clusters in $1.2 < z < 1.8$, found the median concentration index of galaxies in two clusters are higher than that in field. At a glance, our results are similar, but they did not separate quiescent from star-forming galaxies, so the higher medians in the two clusters in their work could be caused by higher fractions of quiescent, early-type galaxies. Since our full sample consists of only HAEs, it is possible that these differences are caused by HAEs starting morphological transformation by building up their bulges \citep[e.g., ][]{Ikeda2022High-resolution1.46}, although it is also likely that there are AGN in our sample that are not detected in X-ray which may have more compact and concentrated morphologies. Meanwhile, \citet{Sazonova2020} found median disturbances (as derived from Gini-M$_{20}$; equivalent to merger statistics) for three clusters are higher than that in field, as also shown by the results in our work (Section \ref{sec:mergers}).

%In Figure \ref{fig:morph-by-sfr-offset}, we find that the median concentration of protocluster HAEs is similar to that of field HAEs in the lowest SFMS offset bin, while in the highest SFMS offset bin protocluster HAEs are slightly more concentrated.
%Concentration is defined as the ratio between $r_{80}$ and $r_{20}$ (see \ref{sec:concentration}). While we do not find significant difference between protocluster HAEs and field HAEs in size, both in $r_\mathrm{Sersic}$ as we presented and in the non-parametrically measured $r_{50}$, we find that at fixed $r_{80}$, protocluster HAEs have smaller $r_{20}$ than field HAEs---i.e., higher concentration. Since the concentrated profiles are more prominent in highly star-forming protocluster galaxies, this may indicate that protocluster starbursts are more confined to the central parts of the galaxies.

We also find that asymmetry and Gini-M$_{20}$ merger statistic increases as the star-formation rate offset increases. Merging galaxies may show an increase in asymmetry and/or concentration depending on the stage of mergers, although many variables affect the observed quantitative morphology, such as mass ratio and gas fractions \citep{Lotz2010a, lotz2010b}. However, the irregularities in the morphology of galaxies in our sample may also be caused by giant in-situ star-forming clumps. Giant clumps can form in a galaxy due to disk instability caused by cold stream of gas \citep[][]{Dekel2009} and then migrate to the center of the galaxy.

\subsection{Disturbed fractions} \label{sec:mergers}

Non-parametric morphology statistics are often used to identify galaxies with merger signatures, as they usually satisfy some criteria of non-parametric statistics, with the goal of estimating merger fractions and merger rates \citep[e.g., ][]{Conselice2003, Lotz2004, Conselice2014}. However, the nature of peculiar galaxies at high-redshift is not clear, with the small apparent size, lower surface brightness, and morphological k-correction. Here we estimate the fraction of disturbed galaxies using the commonly used criteria, while keeping in mind that the cause of the peculiar morphologies may be ambiguous.

Adopting $A > 0.3$ as merger candidates/disturbed galaxies \citep[][]{Conselice2003, Lotz2008, Peth2016, Hine2016}, we find $16 \pm 7\%$ protocluster HAEs are disturbed, compared to $14 \pm 3\%$ field HAEs. Note that $\sim 20 \%$ of objects in our sample have negative $A$, which arises when the local background is high and the galaxy is sufficiently low surface brightness, so we might have missed more disturbed when using this criterion.

As described in Section \ref{sec:morphologies}, Gini-M$_{20}$ merger statistics are calculated by \textsc{Statmorph}. The criterion based on Gini-M$_{20}$ is more sensitive to minor mergers and gas-poor mergers than asymmetry and also more robust to low signal-to-noise ratio \citep{Lotz2004, Lotz2008, Lotz2010a, Peth2016}. Using this criterion, we estimate $39 \pm 8\%$ protocluster HAEs are disturbed. This fraction is higher than that of the field, $26 \pm 4\%$, suggesting enhanced disturbed fraction in protoclusters.

\begin{figure}
    \centering
    \includegraphics{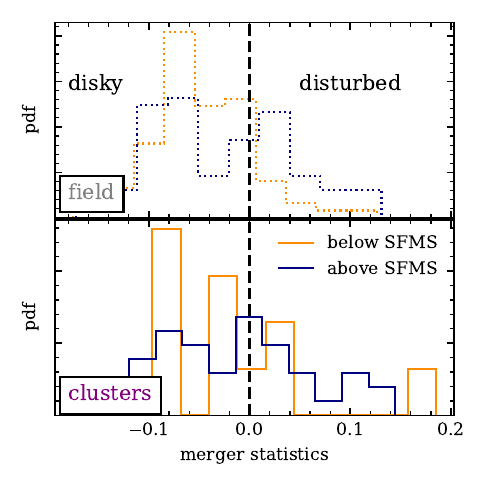}
    \caption{Gini-M$_{20}$ mergers statistic divided into two SFMS offset bins: blue for those above the star-forming main sequence and orange for those below the star-forming main sequence. We find that higher disturbed galaxies fraction in protoclusters. The disturbed fraction is also correlated with SFR both in protoclusters and in field.}
    \label{fig:gini-m20-sfr}
\end{figure}

In Figure \ref{fig:gini-m20-sfr}, we show a histogram of merger statistics derived by Gini-M$_{20}$, color-coded by the position relative to the star-forming main sequence. We find $43 \pm 11 \%$ protocluster and $34 \pm 8\%$ field starbursting HAEs are disturbed based on this criterion. This is quite consistent with the finding from \citet{Stott2013} that $40-50\%$ starburst galaxies at $z=2.23$ from HiZELS have a merger-like morphology, although they only use M$_{20}$ as the criterion of merger candidates.

We should note that our results are based on rest-frame UV view ($\sim 2500 $ Å) of high-redshift star-forming galaxies, which is biased by short-lived massive young star populations and may be severely affected by dust. The disturbed galaxies in our sample are likely galaxies with prominent star-forming clumps instead of mergers. \citet{Wuyts2012} show that high-redshift star-forming galaxies are more clumpy in rest-frame UV than in rest-frame optical and in stellar mass distribution. \citet{Shibuya2016} argued that irregular galaxies at high redshift are not due to mergers, but due to violent disk instabilites. The smoothness index $S$ is sometimes used to distinguish asymmetries due to mergers from clumps \citep{Conselice2008}, i.e., an additional criterion $A > S$ is used. However, we find that \textsc{STATMORPH} implementation of smoothness is not very effective in detecting star-formation clumps in our sample, as $90\%$ of our objects are measured to have $S < 0.06$.

If we assume the disturbed galaxies are due to in-situ clumps, the fact that protocluster HAEs are more clumpy than their field counterparts still remains. In-situ clumps may be caused by streams of cold gas to galactic disks, and continuous streams may sustain the instabilities in the disk \citep{Dekel2009}. Thus, the more clumpy morphologies in protoclusters might point out to different gas inflow patterns in different environments. Nonetheless, our goal is to compare star-forming galaxies in different environments---while the interpretation of these results is still unclear, we find that there are some differences between protocluster and field HAE morphologies. We believe that this is worth investigating in future studies.

\subsection{Rest-frame optical view}
The morphology of galaxies is known to be dependent on the probing wavelength \citep{Conselice2014, Wuyts2011b}. The apparent size might also be affected, as hinted by the discrepancy between the mass-size relations of this work and \citet{VanDerWel2014}, although \citet{Shen2023} also shows that the NUV size of $z\sim2$ galaxies are quite consistent with continuum size.

As a quick check, we performed a stacking analysis following the same procedure as in \ref{sec:stacking} for ground-based Ks-band data, which covers the rest-frame optical regime of our $z\sim2$ HAEs. We use ESO/UltraVISTA Ks-band data \citep{McCracken2012} for our COSMOS sample, both protoclusters (PHzG237 and CC2.2) and the comparison field. Note that while COSMOS field was observed by HST WFC3 in F160W filter for CANDELS, only a small subset of our COSMOS sample is available in F160W. We obtain the half-light radii for stacked protocluster galaxies $r_\mathrm{Sersic} = 2.65^{+1.19}_{-0.80}$ kpc and for stacked field galaxies $r_\mathrm{Sersic} = 2.57^{+0.70}_{-0.63}$ kpc. We do not find environmental difference on rest-frame optical size, agreeing with our rest-frame UV results.

We do not analyze the morphology of individual galaxies with the K-band data due to the large seeing size. Seeing blurs out the detailed morphological features, making it more difficult to detect clumps or asymmetries. As of the writing of this paper, several papers have been published or submitted that utilized JWST data to investigate the rest-frame optical/NIR morphology of $z > 2$ galaxies. For example, \citet{Ferreira2022} classified the morphology of four thousand galaxies at $1.5 < z < 8$, visually and quantitatively. They find that high redshift galaxies can be unambiguously classified into disks, spheroids, or peculiars, with a surprisingly high fraction of regular disk galaxies as far as $z \sim 6$. JWST shows that some galaxies with unclear morphology in HST/F814W turn out to be disk galaxies with bars and/or spiral arms, showing the wavelength dependence of morphology.

%%%%%%
\section{Summary} \label{sec:summary}

We analyzed the rest-frame UV size and morphologies of 122 (436) H$\alpha$-selected star-forming galaxies in protoclusters (field) at $z\sim2$ to investigate whether the structures of star-forming galaxies are affected by their environment at this redshift. The following list summarizes our results.

\begin{enumerate}
    \item The size distributions of protocluster and field HAEs at $z \sim 2$ are similar with typical half-light radius of $\sim2.5$ kpc. At fixed stellar mass, there is no significant difference between HAE in protocluster and in field. Stacking analyses also show similar results. This result suggests that the environment does not significantly affect the size of galaxies during the star-forming phase.
    \item HAE morphologies at $z \sim 2$ are consistent with disk-like star-forming galaxies, but we also find $29\% \pm 4\%$ of all HAEs in our sample have disturbed morphologies.
    %\item Based on Sérsic index of stacked images and non-parametric morphologies, protocluster HAEs are on average have slightly more concentrated light profile than field HAEs. This may indicate that the galaxies in protoclusters are starting to build up their bulges.
    \item Based on Gini-M$_{20}$ merger statistics, we estimate the disturbed fraction is $38 \pm 7 \%$ in protoclusters and $26 \pm 3\%$ in field. We also find that there are twice disturbed galaxies above star-forming main sequence than below it, indicating a correlation between star-formation activity and peculiar morphology. This fact indicates merger-driven starbursts or in-situ star-forming clumps caused by continuous cold gas inflows, both of which may be a function of environment.
\end{enumerate}

The caveat of our study is that our morphological analysis are solely based on HST/ACS F814W data, which probes rest-frame $\sim 2500$ Å wavelength at $z = 2$. This regime is dominated by massive young stars and does not really represent the underlying stellar mass distribution. We also did not consider how dust affects the structures of the galaxies. Despite these caveats, we did find some differences in the rest-frame UV morphologies of HAEs (e.g., point 3), which hints some degree of environmental effects in the apparent morphologies. Wide surveys by JWST, such as COSMOS-Web \citep{Casey2022COSMOS-Web:Survey}, and the upcoming JWST observation of PKS1138 (PropID: 1572, PI: Helmut Dannerbauer, Co-PI: Yusei Koyama) will provide high-resolution rest-frame optical images of high-redshift galaxies, allowing us to see the more complete stellar distribution to study their structures as a function of environment.

\begin{acknowledgements}
This work is based on data collected at the Subaru Telescope, which is operated by the National Astronomical Observatory of Japan. We are honored and grateful for the opportunity of observing the Universe from Maunakea, which has the cultural, historical, and natural significance in Hawaii. A subset of our dataset was provided by David Sobral and the HiZELS team. We also thank Masami Ouchi, Daisuke Iono, and Masatoshi Imanishi for the discussions. Some data were obtained from the NASA/ESA Hubble Space Telescope through Hubble Legacy Archive, which is a collaboration between the Space Telescope Science Institute (STScI/NASA), the Space Telescope European Coordinating Facility (ST-ECF/ESA) and the Canadian Astronomy Data Centre (CADC/NRC/CSA). This work made use of Astropy:\footnote{http://www.astropy.org} a community-developed core Python package and an ecosystem of tools and resources for astronomy (Astropy 2013, 2018, 2022).

This work was financially supported in part by a Grant-in-Aid for the Scientific Research (No.18K13588 and 23H01219) by the Japanese Ministry of Education, Culture, Sports and Science.
\end{acknowledgements}

\bibliography{references}{}
\bibliographystyle{aasjournal}

\end{document}